\documentclass[aps,pra,twocolumn,superscriptaddress,showpacs]{revtex4}
\usepackage{graphicx} 
\usepackage{bm}

\begin{document}

\title{Fractional ac Josephson effect in unconventional
  superconductors}

\author{Hyok-Jon Kwon}
\affiliation{Department of Physics, University of Maryland, College
  Park, Maryland 20742-4111, USA}

\author{K. Sengupta} 
\affiliation{Department of Physics, Yale University,
   New Haven, Connecticut 06520-8120, USA}

\author{Victor M. Yakovenko}
\affiliation{Department of Physics, University of Maryland, College
  Park, Maryland 20742-4111, USA}

\date{\bf cond-mat/0401313, v.1 18 January 2004, v.2 3 June 2004}


\begin{abstract}
  For certain orientations of Josephson junctions between two
  $p_x$-wave or two $d$-wave superconductors, the subgap Andreev bound
  states produce a $4\pi$-periodic relation between the Josephson
  current $I$ and the phase difference $\phi$: $I\propto\sin(\phi/2)$.
  Consequently, the ac Josephson current has the fractional frequency
  $eV/\hbar$, where $V$ is the dc voltage.  In the tunneling limit,
  the Josephson current is proportional to the first power (not
  square) of the electron tunneling amplitude.  Thus, the Josephson
  current between unconventional superconductors is carried by single
  electrons, rather than by Cooper pairs.  The fractional ac Josephson
  effect can be observed experimentally by measuring frequency
  spectrum of microwave radiation from the junction.
\end{abstract} 

\maketitle

\section{Brief history of the ac Josephson effect}

In 1962, Josephson \cite{Josephson} predicted theoretically that if a
dc voltage $V$ is applied to a junction between two superconductors,
ac supercurrent with the frequency $2eV/\hbar$ appears between the
superconductors.  The ac Josephson radiation was first observed
experimentally 40 years ago in Kharkov by Yanson, Svistunov, and
Dmitrenko \cite{Yanson1,Yanson2}.  In Ref.\ \cite{Yanson2}, the
spectrum of microwave radiation from tin junctions was measured, and a
sharp peak in the frequency spectrum at $2eV/\hbar$ was found.  It is
amazing that without any attempt to match impedances of the junction
and waveguide, Dmitrenko and Yanson \cite{Yanson2} found the signal
several hundred times stronger than the noise and the ratio of
linewidth to the Josephson frequency less than $10^{-3}$.  This
discovery was followed by further work in the United States
\cite{Langenberg} and Ukraine \cite{Yanson3}.  The results of these
investigations were summarized in the classic book
\cite{Kulik-Yanson}.  Since then, the ac Josephson radiation has been
observed in many materials in various experimental setups.  For
example, a peak of Josephson radiation was found in Ref.\ 
\cite{Lukens} in indium junctions at the frequency 9 GHz with the
width 36 MHz.  In Ref.\ \cite{Mulller}, a peak of Josephson radiation
was observed around 11 GHz with the width 50 MHz in $\rm
Bi_2Sr_2CaCu_2O_8$ single crystals with the current along the $\bm{c}$
axis perpendicular to the layers.

The theory of the Josephson effect was originally developed for
conventional $s$-wave superconductors.  In this paper, we study
Josephson junctions between unconventional superconductors, such as
$d$-wave cuprates or $p_x$-wave organic superconductors.  We show that
the midgap Andreev states in these materials produce a $4\pi$-periodic
relation between the Josephson current $I$ and the phase difference
$\phi$: $I\propto\sin(\phi/2)$.  Consequently, the ac Josephson
current has the fractional frequency $eV/\hbar$, a half of the
conventional value.  We hope that this effect can observed
experimentally as the corresponding peak in the frequency spectrum of
Josephson radiation from unconventional superconductors, such as
$d$-wave cuprates, in the manner similar to the pioneering experiments
\cite{Yanson1,Yanson2} performed on conventional $s$-wave
superconductors.

\section{Introduction}

In many materials, the symmetry of the superconducting order parameter
is unconventional, i.e.\ not $s$-wave.  In the high-$T_c$ cuprates, it
is the singlet $d_{x^2-y^2}$-wave \cite{vanHarlingen}.  There is
experimental evidence that, in the quasi-one-dimensional (Q1D) organic
superconductors $\rm(TMTSF)_2X$ \cite{TMTSF}, the symmetry is triplet
\cite{Chaikin}, most likely the $p_x$-wave \cite{Lebed}, where the $x$
axis is along the conducting chains.  The unconventional pairing
symmetry typically results in formation of midgap Andreev bound states
on the surfaces of these superconductors.  For $d$-wave cuprate
superconductors, the midgap Andreev states were predicted
theoretically in Ref.\ \cite{Hu} and discovered experimentally as a
zero-bias conductance peak in tunneling between normal metals and
superconductors (see review \cite{Tanaka-review}).  For the Q1D
organic superconductors, the midgap states were theoretically
predicted to exist at the edges perpendicular to the chains
\cite{Ours1,Tanuma}.  When two unconventional superconductors are
joined together in a Josephson junction, their Andreev surface states
hybridize to form Andreev bound states in the junction.  These states
are important for the Josephson current.  Andreev bound states in
high-$T_c$ junctions were reviewed in Ref.\ \cite{Wendin-review}.  The
Josephson effect between two Q1D $p$-wave superconductors was studied
in Refs.\ \cite{Tanaka-pp,Vaccarella}.

In the present paper, we predict a new effect for Josephson junctions
between unconventional (nonchiral) superconductors, which we call the
fractional ac Josephson effect.  Suppose both superconductors forming
a Josephson junction have surface midgap states originally.  This is
the case for the butt-to-butt junction between two $p_x$-wave Q1D
superconductors, as shown in Fig.\ \ref{fig:JJ}a, and for the
$45^\circ/45^\circ$ in-plane junction between two $d$-wave
superconductors, as shown in Fig.\ \ref{fig:d-wave}a.  (The two angles
indicate the orientation of the junction line relative to the $\bm{b}$
axes of each $d_{x^2-y^2}$ superconductor.)  We predict that the
contribution of the hybridized Andreev bound states produces a
$4\pi$-periodic relation between the supercurrent $I$ and the
superconducting phase difference $\phi$: $I\propto\sin(\phi/2)$
\cite{2phi}.  Consequently, the ac Josephson effect has the frequency
$eV/\hbar$, where $e$ is the electron charge, $V$ is the applied dc
voltage, and $\hbar$ is the Planck constant.  The predicted frequency
is a half of the conventional Josephson frequency $2eV/\hbar$
originating from the conventional Josephson relation
$I\propto\sin{\phi}$ with the period of $2\pi$.  Qualitatively, the
predicted effect can be understood as follows.  The Josephson current
across the two unconventional superconductors is carried by tunneling
of \emph{single electrons} (rather than Cooper pairs) between the two
resonant midgap states.  Thus, the Cooper pair charge $2e$ is replaced
the single charge $e$ in the expression for the Josephson frequency.
This interpretation is also supported by the finding that, in the
tunneling limit, the Josephson current is proportional to the first
power (not square) of the electron tunneling amplitude
\cite{Tanaka2,Riedel,Barash}.  Possibilities for experimental
observation of the fractional ac Josephson effect are discussed in
Sec.\ \ref{sec:experiment}.

The predicted current-phase relation $I\propto\sin(\phi/2)$ is quite
radical, because every textbook on superconductivity says that the
Josephson current must be a $2\pi$-periodic function of $\phi$
\cite{2phi}.  To our knowledge, the only paper that discussed the
$4\pi$-periodic Josephson effect is Ref.\ \cite{Kitaev} by Kitaev.  He
considered a highly idealized model of spinless fermions on a
one-dimensional (1D) lattice with superconducting pairing on the
neighboring sites.  The pairing potential in this case has the
$p_x$-wave symmetry, and midgap states exist at the ends of the chain.
They are described by the Majorana fermions, which Kitaev proposed to
use for nonvolatile memory in quantum computing.  He found that, when
two such superconductors are brought in contact, the system is
$4\pi$-periodic in the phase difference between the superconductors.
Our results are in agreement with his work.  However, we formulate the
problem as an experimentally realistic Josephson effect between known
superconducting materials.

\section{The basics}

In this paper, we consider singlet pairing and triplet pairing with
the spin polarization vector $\bm{n}$ having a uniform,
momentum-independent orientation \cite{Chaikin,Lebed}.  If the spin
quantization axis $z$ is selected along $\bm{n}$, then the Cooper
pairing takes place between electrons with the opposite $z$-axis spin
projections $\sigma$ and $\bar\sigma$: $\langle\hat
c_\sigma(\bm{k})\hat c_{\bar\sigma}(-\bm{k})\rangle
\propto\Delta_\sigma(\bm{k})$, where $\hat c_\sigma(\bm{k})$ is the
annihilation operator of an electron with momentum $\bm{k}$ and spin
$\sigma$.  The pairing potential has the symmetry
$\Delta_\sigma(\bm{k})=\mp\Delta_{\bar\sigma}(\bm{k})
=\pm\Delta_\sigma(-\bm{k})$, where the upper and lower signs
correspond to the singlet and triplet cases.

We select the coordinate axis $x$ perpendicular to the Josephson
junction plane.  We assume that the interface between the two
superconductors is smooth enough, so that the electron momentum
component $k_y$ parallel to the junction plane is a conserved good
quantum number.

Electron states in a superconductor are described by the Bogoliubov
operators $\hat\gamma$, which are related to the electron operators
$\hat c$ by the following equations \cite{Zagoskin}
\begin{eqnarray}
  && \hat{\gamma}_{n\sigma k_y} = \int dx\,
  [u_{n\sigma k_y}^*(x) \, \hat{c}_{\sigma k_y}(x)
  +v_{n\sigma k_y}^*(x) \, \hat{c}_{\bar\sigma\bar k_y}^\dag(x)],
\label{gamma_n} \\
  && \hat c_{\sigma k_y}(x) = \sum_{n} 
  [u_{n\sigma k_y}(x) \, \hat\gamma_{n\sigma k_y}  
  + v_{n\bar\sigma\bar k_y}^*(x) \,
  \hat\gamma_{n\bar\sigma\bar k_y}^\dag],
\label{canon}
\end{eqnarray}
where $\bar k_y=-k_y$, and $n$ is the quantum number of the Bogoliubov
eigenstates.  The two-components vectors $\psi_{n\sigma
  k_y}(x)=[u_{n\sigma k_y}(x),v_{n\sigma k_y}(x)]$ are the eigenstates
of the Bogoliubov-de Gennes (BdG) equation with the eigenenergies
$E_{n\sigma k_y}$:
\begin{equation}
  \left(\begin{array}{cc} 
  \varepsilon_{k_y}(\hat k_x)+U(x) & 
  \hat\Delta_{\sigma k_y}(x,\hat k_x) \\ 
  \hat\Delta_{\sigma k_y}^\dag(x,\hat k_x) & 
  -\varepsilon_{k_y}(\hat k_x)-U(x)
  \end{array}\right) 
  \psi_{n} = E_{n} \psi_{n},
\label{eq:BdG}
\end{equation}
where $\hat k_x=-i\partial_x$ is the $x$ component of the electron
momentum operator, and $U(x)$ is a potential.  In Eq.\ (\ref{eq:BdG})
and below, we often omit the indices $\sigma$ and $k_y$ to shorten
notation where it does not cause confusion.

\section{Junctions between quasi-one-dimensional superconductors}
\label{sec:Q1D}

In this section, we consider junctions between two Q1D
superconductors, such as organic superconductors $\rm(TMTSF)_2X$, with
the chains along the $x$ axis, as shown in Fig.\ \ref{fig:JJ}a.  For a
Q1D conductor, the electron energy dispersion in Eq.\ (\ref{eq:BdG})
can be written as $\varepsilon=\hbar^2\hat
k_x^2/2m-2t_b\cos(bk_y)-\mu$, where $m$ is an effective mass, $\mu$ is
the chemical potential, $b$ and $t_b$ are the distance and the
tunneling amplitude between the chains.  The superconducting pairing
potentials in the $s$- and $p_x$-wave cases have the forms
\begin{eqnarray}
  \hat\Delta_{\sigma k_y}(x,\hat k_x) &=& \left\{ 
  \begin{array}{cc}
  \sigma\Delta_\beta, & \mbox{$s$-wave}, \\
  \Delta_\beta\,\hat k_x/k_F , & \mbox{$p_x$-wave},
  \end{array} \right.
\label{hat-Delta}
\end{eqnarray}
where $\hbar k_F=\sqrt{2m\mu}$ is the Fermi momentum, and $\sigma$ is
treated as $+$ for $\uparrow$ and $-$ for $\downarrow$.  The index
$\beta=R,L$ labels the right ($x>0$) and left ($x<0$) sides of the
junction, and $\Delta_\beta$ acquires a phase difference $\phi$ across
the junction:
\begin{equation}
  \Delta_L=\Delta_0, \qquad
  \Delta_R=\Delta_0e^{i\phi}.
\label{gapfn}
\end{equation}
The potential $U(x)=U_0\delta(x)$ in Eq.\ (\ref{eq:BdG}) represents
the junction barrier located at $x=0$.  Integrating Eq.\ 
(\ref{eq:BdG}) over $x$ from --0 to +0, we find the boundary
conditions at $x=0$:
\begin{eqnarray}
  && \psi_L=\psi_R,\quad
  \partial_x\psi_R - \partial_x\psi_L = k_F Z\,\psi(0),
\label{Bdy} \\
  && Z=2mU_0/\hbar^2k_F, \quad D=4/(Z^2+4),
\label{ZDG}
\end{eqnarray}
where $D$ is the transmission coefficient of the barrier.

\subsection{Andreev bound states}
\label{sec:bound}

A general solution of Eq.\ (\ref{eq:BdG}) is a superposition of the
terms with the momenta close to $\alpha k_F$, where the index
$\alpha=\pm$ labels the right- and left-moving electrons:
\begin{equation}
  \psi_{\beta\sigma} = {e^{\beta\kappa x}} \left[ 
  A_\beta \left(
  \begin{array}{c} u_{\beta\sigma+} \\ v_{\beta\sigma+} \end{array}
  \right) e^{i\tilde k_Fx} 
  + B_\beta \left( 
  \begin{array}{c} u_{\beta\sigma-} \\ v_{\beta\sigma-} \end{array} 
  \right) e^{-i\tilde k_Fx} \right].
\label{L0s}
\end{equation}
Here $\beta=\mp$ for $R$ and $L$.  Eq.\ (\ref{L0s}) describes a subgap
bound state with an energy $|E|<\Delta_0$, which is localized at the
junction and decays exponentially in $x$ within the length $1/\kappa$.
The coefficients $(u_{\beta\sigma\alpha},v_{\beta\sigma\alpha})$ in
Eq.\ (\ref{L0s}) are determined by substituting the right- and
left-moving terms separately into Eq.\ (\ref{eq:BdG}) for $x\neq0$,
where $U(x)=0$.  In the limit $k_F\gg\kappa$, we find
\begin{equation}
  {u_{\beta\sigma\alpha} \over v_{\beta\sigma\alpha}}=
  {\Delta_{\beta\sigma\alpha} \over {E+i\alpha\beta\hbar\kappa v_F}},
  \quad  \kappa={\sqrt{\Delta_0^2-|E|^2} \over \hbar v_F}, 
\label{kappa-eta}
\end{equation}
where $v_F=\hbar k_F/m$ is the Fermi velocity, and
$\Delta_{\beta\sigma\alpha}$ is equal to $\sigma\Delta_\beta$ for
$s$-wave and to $\alpha\Delta_\beta$ for $p_x$-wave, with
$\Delta_\beta$ given by Eq.\ (\ref{gapfn}).  The $k_y$-dependent Fermi
momentum $\hbar\tilde k_F=\hbar k_F+2t_b\cos(bk_y)/v_F$ in Eq.\ 
(\ref{L0s}) eliminates the dispersion in $k_y$ from the BdG equation.

Substituting Eq.\ (\ref{L0s}) into the boundary conditions
(\ref{Bdy}), we obtain the linear homogeneous equations for the
coefficients $A_\beta$ and $B_\beta$.  The compatibility condition for
these equations gives an equation for the energies of the Andreev
bound states.  There are two subgap states with the energies
$E_a=aE_0(\phi)$ labeled by the index $a=\pm$:
\begin{eqnarray}
  E_0^{(s)}(\phi) &=&- \Delta_0 \sqrt{1-D\sin^2(\phi/2)},
  \; \mbox{$s$-$s$ junction},
\label{E_s} \\
  E_0^{(p)}(\phi) &=& -\Delta_0 \sqrt{D}\cos(\phi/2),
  \quad \mbox{$p_x$-$p_x$ junction}.
\label{E_p}
\end{eqnarray}

\begin{figure}[t] 
\includegraphics[width=\linewidth,angle=0]{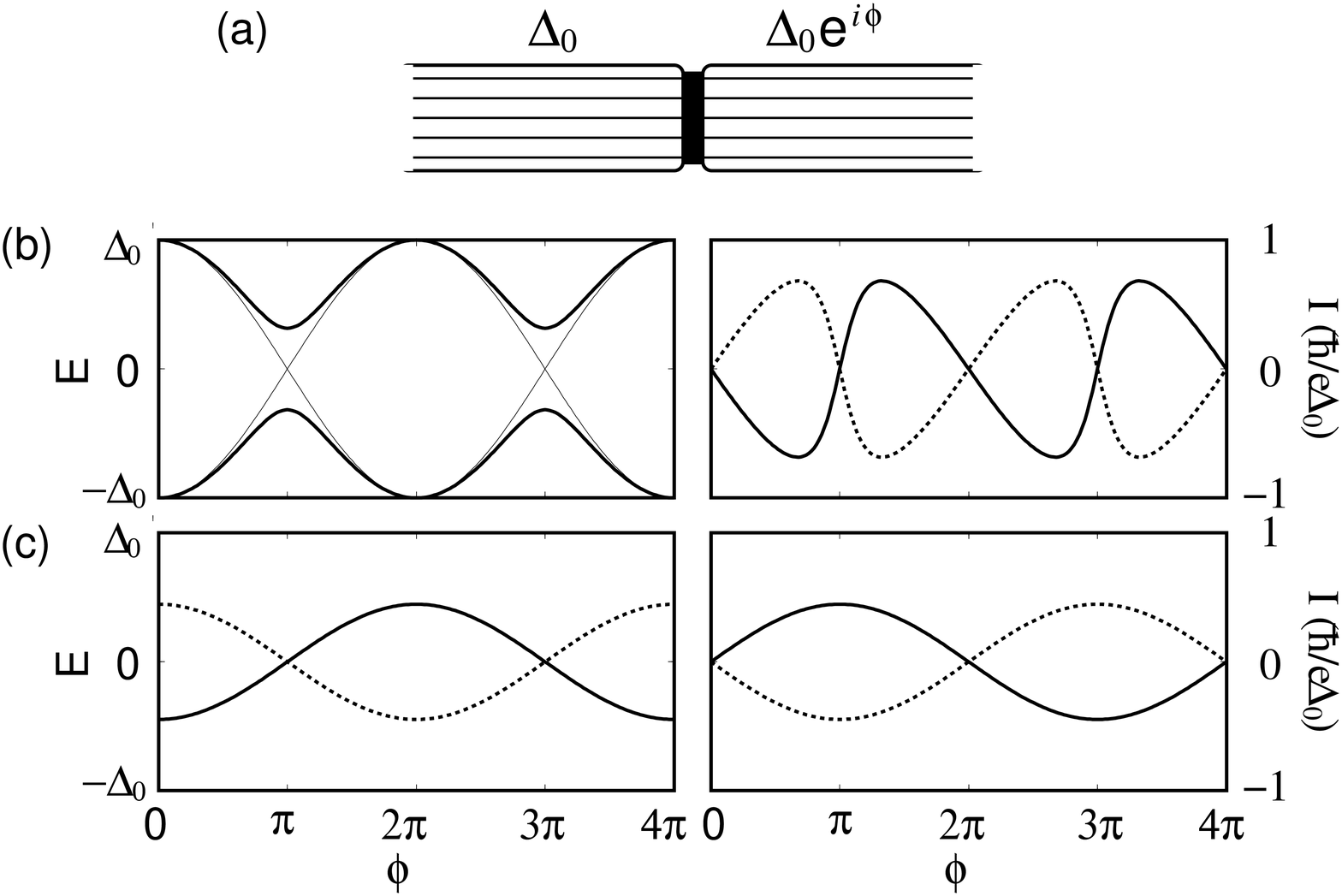} 
\caption{
  (a) Josephson junction between two Q1D $p_x$-wave superconductors.
  (b) The energies (left panel) and the currents (right panel) of the
  subgap states in the $s$-$s$ junction as functions of the phase
  difference $\phi$ for $D=1$ (thin lines) and $D=0.9$ (thick lines).
  (c) The same as (b) for the $p_x$-$p_x$ junction at $D=0.2$.}
\label{fig:JJ}
\end{figure}

The energies (\ref{E_s}) and (\ref{E_p}) are plotted as functions of
$\phi$ in the left panels (b) and (c) of Fig.\ \ref{fig:JJ}.  Without
barrier ($D=1$), the spectra of the $s$-$s$ and $p_x$-$p_x$ junctions
are the same and consist of two crossing curves $E=\mp
\Delta_0\cos(\phi/2)$, shown by the thin lines in the left panel of
Fig.\ \ref{fig:JJ}b.  A non-zero barrier ($D<1$) changes the energies
of the Andreev bound states in the $s$-$s$ and $p_x$-$p_x$ junctions
in different ways.  In the $s$-$s$ case, the two energy levels repel
near $\phi=\pi$ and form two separated $2\pi$-periodic branches shown
by the thick lines in the left panel of Fig.\ \ref{fig:JJ}b
\cite{Zagoskin,Furusaki99}.  In contrast, in the $p_x$-$p_x$ case, the
two energy levels continue to cross at $\phi=\pi$, and they separate
from the continuum of states above $+\Delta_0$ and below $-\Delta_0$,
as show in the left panel of Fig.\ \ref{fig:JJ}c.  The absence of
energy levels repulsion indicates that there is no matrix element
between these levels at $\phi=\pi$ in the $p_x$-$p_x$ case, unlike in
the $s$-$s$ case.

As shown in Sec.\ \ref{sec:d-wave}, the $45^\circ/45^\circ$ junction
between two $d$-wave superconductors is mathematically equivalent to
the $p_x$-$p_x$ junction.  Eq.\ (\ref{E_p}) was derived for the
$45^\circ/45^\circ$ junction in Ref.\ \cite{Tanaka1,Riedel,Barash}.

\subsection{dc Josephson effect in thermodynamic equilibrium}
\label{sec:dc}

It is well known \cite{Zagoskin,A-B} that the current carried by a
quasiparticle state $a$ is
\begin{equation}
  I_a={2e \over \hbar}\,{\partial E_a \over \partial\phi}.
\label{eq:I_a}  
\end{equation}
The two subgap states carry opposite currents, which are plotted vs.\ 
$\phi$ in the right panels (b) and (c) of Fig.\ \ref{fig:JJ} for the
$s$-$s$ and $p_x$-$p_x$ junctions.  In thermodynamic equilibrium, the
total current is determined by the Fermi occupation numbers $f_a$ of
the states at a temperature $T$:
\begin{equation}
  I= {2e\over\hbar}\sum_{a=\pm}
  {\partial E_a\over\partial\phi}\,f_a=
  -{2e\over\hbar}{\partial E_0\over\partial\phi}
  \tanh\left({E_0\over2T}\right).
\label{thermal}
\end{equation}
For the $s$-$s$ junction, substituting Eq.\ (\ref{E_s}) into Eq.\ 
(\ref{thermal}), we recover the Ambegaokar-Baratoff formula \cite{AB}
in the tunneling limit $D\ll1$
\begin{equation}
  I_s \approx D\sin\phi\,{e\Delta_0\over2\hbar} 
  \tanh\left({\Delta_0\over2T}\right)
  =\sin\phi\,{\pi\Delta_0\over2eR}
  \tanh\left({\Delta_0\over2T}\right)
\label{I_s}
\end{equation}  
and the Kulik-Omelyanchuk formula \cite{KO} in the transparent limit
$D\to1$  
\begin{equation}
I_s \approx \sin\left({\phi\over2}\right)
  {e\Delta_0\over\hbar} 
  \tanh\left({\Delta_0\cos(\phi/2)\over2T}\right).
\label{I_sKO}
\end{equation}  
Taking into account that the total current is proportional to the
number $N$ of conducting channels in the junction (e.g.\ the number of
chains), we have replaced the transmission coefficient $D$ in Eq.\ 
(\ref{I_s}) by the junction resistance $R=h/2Ne^2D$ in the normal
state.

Substituting Eq.\ (\ref{E_p}) into Eq.\ (\ref{thermal}), we find the
Josephson current in the $p_x$-$p_x$ junction in thermodynamic
equilibrium:
\begin{eqnarray}
  && I_p = \sqrt{D}\sin\left({\phi\over2}\right)
  {e\Delta_0\over\hbar} 
  \tanh\left({\Delta_0\sqrt{D}\cos(\phi/2)\over2T}\right)
\nonumber \\
  && = \sin\left({\phi\over2}\right)
  {\pi\Delta_0\over\sqrt{D}eR}
  \tanh\left({\Delta_0\sqrt{D}\cos(\phi/2)\over2T}\right).
\label{I_p}
\end{eqnarray}

\begin{figure}[t] 
\includegraphics[width=0.8\linewidth,angle=0]{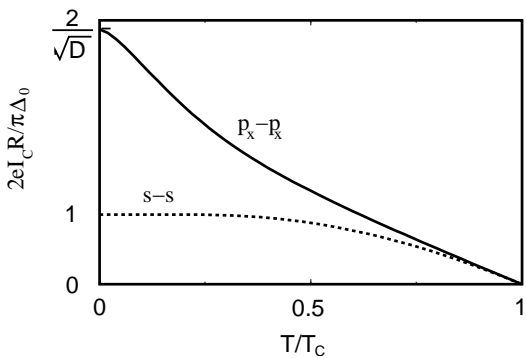} 
\caption{
  Critical currents of the $s$-$s$ (dashed line) and $p_x$-$p_x$
  (solid line) Josephson junctions as functions of temperature for
  $D=0.3$.}
\label{fig:Ic} 
\end{figure}

The temperature dependences of the critical currents for the $s$-$s$
and $p_x$-$p_x$ junctions are shown in Fig.\ \ref{fig:Ic}.  They are
obtained from Eqs.\ (\ref{I_s}) and (\ref{I_p}) assuming the BCS
temperature dependence for $\Delta_0$.  In the vicinity of $T_c$,
$I_p$ and $I_s$ have the same behavior.  With the decrease of
temperature, $I_s$ quickly saturates to a constant value, because, for
$D\ll1$, $E_a^{(s)}\approx\mp\Delta_0$ (\ref{E_s}), thus, for
$T\alt\Delta_0$, the upper subgap state is empty and the lower one is
completely filled.  In contrast, $I_p$ rapidly increases with
decreasing temperature as $1/T$ and saturates to a value enhanced by
the factor $2/\sqrt{D}$ relative to the Ambegaokar-Baratoff formula
(\ref{E_s}) at $T=0$.  This is a consequence of two effects.  As Eqs.\ 
(\ref{I_s}) and (\ref{I_p}) show, $I_s\propto D$ and $I_p\propto
\sqrt{D}$, thus $I_p\gg I_s$ in the tunneling limit $D\ll1$.  At the
same time, the energy splitting between the two subgap states in the
$p_x$-$p_x$ junction is small compared to the gap:
$E_0^{(p)}\propto\sqrt{D}\Delta_0\ll\Delta_0$.  Thus, for
$\sqrt{D}\Delta_0\alt T\alt\Delta_0$, the two subgap states are almost
equally populated, so the critical current has the $1/T$ temperature
dependence analogous to the Curie spin susceptibility.

Eq.\ (\ref{I_p}) was derived analytically for the $45^\circ/45^\circ$
junction between two $d$-wave superconductors in Refs.\ 
\cite{Riedel,Tanaka2}, and a similar result was calculated numerically
for the $p_x$-$p_x$ junction in Ref.\ \cite{Tanaka-pp,Vaccarella}.
Notice that Eq.\ (\ref{I_p}) gives the Josephson current $I_p(\phi)$
that is a $2\pi$-periodic functions of $\phi$, both for $T=0$ and
$T\neq0$.  This is a consequence of the thermodynamic equilibrium
assumption.  At $T=0$, this assumption implies that the subgap state
with the lower energy is occupied, and the one with the higher energy
is empty.  As one can see in Fig.\ \ref{fig:JJ}, the \emph{lower}
energy is always a $2\pi$-periodic functions of $\phi$.  The
assumption of thermodynamic equilibrium was explicitly made in Ref.\ 
\cite{Riedel} and was implicitly invoked in Refs.\ 
\cite{Tanaka2,Tanaka-pp,Vaccarella} by using the Matsubara diagram
technique.  In Ref.\ \cite{Arie}, temperature dependence of the
Josephson critical current was measured in the YBCO ramp-edge
junctions with different crystal angles and was found to be
qualitatively consistent with the upper curve in Fig.\ \ref{fig:Ic}.

\subsection{Dynamical fractional ac Josephson effect}

The calculations of the previous section apply in the static case,
where a given phase difference $\phi$ is maintained for an infinitely
long time, so the occupation numbers of the subgap states have enough
time to relax to thermodynamic equilibrium.  Now let us consider the
opposite, dynamical limit.  Suppose a small voltage $eV\ll\Delta_0$ is
applied to the junction, so the phase difference acquires dependence
on time $t$: $\phi(t)=2eVt/\hbar$.  In this case, the state of the
system is determined dynamically starting from the initial conditions.
Let us consider the $p_x$-$p_x$ junction at $T=0$ in the initial state
$\phi=0$, where the two subgap states (\ref{E_p}) with the energies
$\pm E_0$ are, correspondingly, occupied and empty.  If $\phi(t)$
changes sufficiently slowly (adiabatically), the occupation numbers of
the subgap states do not change.  In other words, the states shown by
the solid and dotted lines in Fig.\ \ref{fig:JJ}(c) remains,
correspondingly, occupied and empty.  The occupied state (\ref{E_p})
produces the current (\ref{eq:I_a}):
\begin{equation}
  I_p(t)
  ={\sqrt{D}e\Delta_0\over\hbar}\sin\left({\phi(t)\over2}\right)
  ={\sqrt{D}e\Delta_0\over\hbar}\sin\left({eVt\over\hbar}\right).
\label{I_t}
\end{equation}
The frequency of the ac current (\ref{I_t}) is $eV/\hbar$, a half of
the conventional Josephson frequency $2eV/\hbar$.  The fractional
frequency can be traced to the fact that the energies Eq.\ (\ref{E_p})
and the corresponding wave functions have the period $4\pi$ in $\phi$,
rather than conventional $2\pi$.  Although at $\phi=2\pi$ the spectrum
in the left panel of Fig.\ \ref{fig:JJ}(c) is the same as at $\phi=0$,
the occupation numbers are different: The lower state is empty and the
upper state is occupied.  Only at $\phi=4\pi$ the occupation numbers
are the same as at $\phi=0$.

The $4\pi$ periodicity is the consequence of the energy levels
crossing at $\phi=\pi$.  (In contrast, in the $s$-wave case, the
levels repel at $\phi=\pi$ in Fig.\ \ref{fig:JJ}(b), thus the energy
curves are $2\pi$-periodic.)  As discussed at the end of Sec.\ 
\ref{sec:bound}, there is no matrix element between the crossing
energy levels at $\phi=\pi$.  Thus, there are no transitions between
them, so the occupation numbers of the solid and dotted curves in
Fig.\ \ref{fig:JJ}(c) are preserved.  In order to show this more
formally, we can write a general solution of the time-dependent BdG
equation as a superposition of the two subgap states with the
time-dependent $\phi(t)$: $\psi(t)=\sum_aC_a(t)\,\psi_a[\phi(t)]$.
The matrix element of transitions between the states is proportional
to $\dot\phi\langle\psi_+|\partial_\phi\psi_-\rangle=\dot\phi
\langle\psi_+|\partial_\phi\hat{H}|\psi_-\rangle/(E_--E_+)$.  We found
that it is zero in the $p_x$-wave case, thus there are no transitions,
and the initial occupation numbers of the subgap states at $\phi=0$
are preserved dynamically.

As one can see in Fig.\ \ref{fig:JJ}(c), the system is not in the
ground state when $\pi<\phi<3\pi$, because the upper energy level is
occupied and the lower one is empty.  In principle, the system might
be able to relax to the ground state by emitting a phonon or a photon.
At present time, we do not have an explicit estimate for such
inelastic relaxation time, but we expect that it is quite long.  (The
other papers \cite{Riedel,Tanaka2,Tanaka-pp,Vaccarella} that
\emph{assume} thermodynamic equilibrium for each value of the phase
$\phi$ do not have an estimate of the relaxation time either.)  To
observe the predicted ac Josephson effect with the fractional
frequency $eV/\hbar$, the period of Josephson oscillations should be
set shorter than the inelastic relaxation time, but not too short, so
that the time evolution of the BdG equation can be treated
adiabatically.  Controlled nonequilibrium population of the upper
Andreev bound state was recently achieved experimentally in an
$s$-wave Josephson junction in Ref.\ \cite{Baselmans}.

Eq.\ (\ref{I_t}) can be generalized to the case where initially the
two subgap states are populated thermally at $\phi=0$, and these
occupation numbers are preserved by dynamical evolution
\begin{eqnarray}
  I_p(t) &=& {2e\over\hbar}\sum_{a}
  {\partial E_a[\phi(t)] \over \partial\phi}\, f[E_a(\phi=0)]
\label{Idyn} \\
  &=& \sin\left({eVt\over\hbar}\right)
  {\pi\Delta_0\over\sqrt{D}eR}
  \tanh\left({\Delta_0\sqrt{D}\over2T}\right).
\label{I_t-T}
\end{eqnarray}
Notice that the periodicities of the dynamical equation (\ref{I_t-T})
and the thermodynamic Eq.\ (\ref{I_p}) are different.  The latter
equation assumes that the occupation numbers of the subgap states are
in instantaneous thermal equilibrium for each $\phi$.

\subsection{Tunneling Hamiltonian approach}

In the infinite barrier limit $D\to0$, the energies $\pm E_0^{(p)}$ of
the two subgap states (\ref{E_p}) degenerate to zero, i.e.\ they
become midgap states.  The wave functions (\ref{L0s}) simplify as
follows:
\begin{eqnarray}
  \psi_{\pm0} &=& {\psi_{L0}(x)\mp\psi_{R0}(x)\over\sqrt{2}}, 
\label{pm0} \\
  \psi_{L0} &=& \sqrt{2\kappa}\,\sin(k_Fx)\,e^{\kappa x}
  \left(\begin{array}{c} 1 \\ i
  \end{array}\right)\theta(-x),
\label{L0} \\
  \psi_{R0} &=& \sqrt{2\kappa}\,\sin(k_Fx)\,e^{-\kappa x}
  \left(\begin{array}{c} e^{i\phi/2} \\ -ie^{-i\phi/2}
 \end{array}\right)\theta(x).
\label{R0} 
\end{eqnarray}
Since at $D=0$ the Josephson junction consists of two semi-infinite
uncoupled $p_x$-wave superconductors, $\psi_{L0}$ and $\psi_{R0}$ are
the wave functions of the surface midgap states \cite{Ours1} belonging
to the left and right superconductors.  Let us examine the properties
of the midgap states in more detail.

If $(u,v)$ is an eigenvector of Eq.\ (\ref{eq:BdG}) with an eigenvalue
$E_n$, then $(-v^*,u^*)$ for $s$-wave and $(v^*,u^*)$ for $p$-wave are
the eigenvectors with the energy $E_{\bar n}=-E_n$.  It follows from
these relations and Eq.\ (\ref{gamma_n}) that $\hat{\gamma}_{\bar
  n\bar\sigma\bar k_y}=C\hat{\gamma}_{n\sigma k_y}^\dag$ with $|C|=1$.
Notice that in the $s$-wave case, because $(u,v)$ and $(-v^*,u^*)$ are
orthogonal for any $u$ and $v$, the states $n$ and $\bar n$ are always
different.  However, in the $p$-wave case, the vectors $(u,v)$ and
$(v^*,u^*)$ may be proportional, in which case they describe the same
state with $E=0$.  The states (\ref{L0}) and (\ref{R0}) indeed have
this property:
\begin{equation}
  v_{L0} = iu_{L0}^*, \qquad v_{R0} = -iu_{R0}^*.
\label{uv0}
\end{equation}
Substituting Eq.\ (\ref{uv0}) into Eq.\ (\ref{gamma_n}), we find the
Bogoliubov operators of the left and right midgap states
\begin{equation}
  \hat{\gamma}^\dag_{L0\sigma k_y}
  =i\hat{\gamma}_{L0\bar\sigma\bar k_y},
  \quad
  \hat{\gamma}^\dag_{R0\sigma k_y}
  =-i\hat{\gamma}_{R0\bar\sigma\bar k_y}.
\label{conjugate}
\end{equation}
Operators (\ref{conjugate}) correspond to the Majorana fermions
discussed in Ref.\ \cite{Kitaev}.  In the presence of a midgap state,
the sum over $n$ in Eq.\ (\ref{canon}) should be understood as
$\sum_{n>0}+(1/2)\sum_{n=0}$, where we identify the second term as the
projection ${\cal P}\hat c$ of the electron operator onto the midgap
state.  Using Eqs.\ (\ref{uv0}), (\ref{conjugate}), and (\ref{canon}),
we find
\begin{equation}
  {\cal P}\hat c_{\sigma k_y}(x) 
  = u_0(x)\hat\gamma_{0\sigma k_y}
  = v_0^*(x)\hat\gamma_{0\bar\sigma\bar k_y}^\dag.
\label{c_0}
\end{equation}

Let us consider two semi-infinite $p_x$-wave superconductors on a 1D
lattice with the spacing $l$, one occupying $x\le\bar l=-l$ and
another $x\ge l$.  They are coupled by the tunneling matrix element
$\tau$ between the sites $\bar l$ and $l$:
\begin{equation}
  \hat H_\tau = \tau \sum_{\sigma k_y} 
  [ \hat{c}^\dag_{L\sigma k_y}(\bar l)\,\hat{c}_{R\sigma k_y}(l) + 
  \hat{c}^\dag_{R\sigma k_y}(l)\,\hat{c}_{L\sigma k_y}(\bar l) ]. 
\label{HT}
\end{equation}
In the absence of coupling ($\tau=0$), the subgap wave functions of
each superconductor are given by Eqs.\ (\ref{L0}) and (\ref{R0}).
Using Eqs.\ (\ref{c_0}), (\ref{uv0}), (\ref{L0}), and (\ref{R0}), the
tunneling Hamiltonian projected onto the basis of midgap states is
\begin{eqnarray}
  && {\cal P}\hat H_\tau = \tau \,
  [u_{L0}^*(\bar l) u_{R0}(l) + {\rm c.c.}] \,
  (\hat\gamma_{L0\uparrow}^\dag \hat\gamma_{R0\uparrow}
  + {\rm H.c.})
\nonumber \\
  && = \Delta_0\sqrt{D}\, \cos(\phi/2)\,
  (\hat\gamma_{L0\uparrow}^\dag \hat\gamma_{R0\uparrow}
  + \hat\gamma_{R0\uparrow}^\dag \hat\gamma_{L0\uparrow}),
\label{PHT}
\end{eqnarray}
where $\sqrt{D}=4\tau\sin^2k_Fl/\hbar v_F$ is the transmission
amplitude, and we omitted summation over the diagonal index $k_y$.
Notice that Eq.\ (\ref{PHT}) is $4\pi$-periodic in $\phi$
\cite{Kitaev}.

Hamiltonian (\ref{PHT}) operates between the two degenerate states of
the system related by annihilation of the Bogoliubov quasiparticle in
the right midgap state and its creation in the left midgap state.  In
this basis, Hamiltonian (\ref{PHT}) can be written as a $2\times2$
matrix
\begin{equation}
  {\cal P}\hat H_\tau = \Delta_0\sqrt{D}\cos(\phi/2)
  \left( \begin{array}{cc} 0 & 1 \\ 1 & 0 \end{array} \right). 
\label{two-level}
\end{equation}
The eigenvectors of Hamiltonian (\ref{two-level}) are $(1,\mp1)$,
i.e.\ the antisymmetric and symmetric combinations of the right and
left midgap states given in Eq.\ (\ref{pm0}).  Their eigenenergies are
$E_\pm(\phi)=\mp\Delta_0\sqrt{D}\cos(\phi/2)$, in agreement with Eq.\ 
(\ref{E_p}).  The tunneling current operator is obtained by
differentiating Eqs.\ (\ref{PHT}) or (\ref{two-level}) with respect to
$\phi$.  Because $\phi$ appears only in the prefactor, the operator
structures of the current operator and the Hamiltonian are the same,
so they are diagonal in the same basis.  Thus, the energy eigenstates
are simultaneously the eigenstates of the current operator with the
eigenvalues
\begin{equation}
  I_\pm=\pm{\sqrt{D}e\Delta_0\over\hbar}
  \sin\left({\phi\over2}\right),
\label{I_pm}
\end{equation}
in agreement with Eq.\ (\ref{I_t}).  The same basis $(1,\mp1)$
diagonalizes Hamiltonian (\ref{two-level}) even when a voltage $V$ is
applied and the phase $\phi$ is time-dependent.  Then the initially
populated eigenstate with the lower energy produces the current
$I_p=\sqrt{D}(e\Delta_0/\hbar)\sin(eVt/\hbar)$ with the fractional
Josephson frequency $eV/\hbar$, in agreement with Eq.\ (\ref{I_t}).

\subsection{Josephson current carried by single electrons, 
  rather than Cooper pairs}
\label{sec:single}

In the tunneling limit, the transmission coefficient $D$ is
proportional to the square of the electron tunneling amplitude $\tau$:
$D\propto\tau^2$.  Eqs.\ (\ref{I_t}) and (\ref{I_pm}) show that the
Josephson current in the $p_x$-$p_x$ junction is proportional to the
first power of the electron tunneling amplitude $\tau$.  This is in
contrast to the $s$-$s$ junction, where the Josephson current
(\ref{I_s}) is proportional to $\tau^2$.  This difference results in
the big ratio $I_p/I_s=2/\sqrt{D}$ between the critical currents at
$T=0$ in the $p_x$- and $s$-wave cases, as shown in Fig.\ \ref{fig:Ic}
and discussed in Sec.\ \ref{sec:dc}.  The reason for the different
powers of $\tau$ is the following.  In the $p_x$-wave case, the
transfer of just one electron between the degenerate left and right
midgap states is a real (nonvirtual) process.  Thus, the eigenenergies
are determined from the secular equation (\ref{two-level}) already in
the first order of $\tau$.  In the $s$-wave case, there are no midgap
states, so the transferred electron is taken from below the gap and
placed above the gap, at the energy cost $2\Delta_0$.  Thus, the
transfer of a single electron is a virtual (not real) process.  It
must be followed by the transfer of another electron, so that the pair
of electrons is absorbed into the condensate.  This gives the current
proportional to $\tau^2$.

This picture implies that the Josephson supercurrent across the
interface is carried by single electrons in the $p_x$-$p_x$ junction
and by Cooper pairs in the $s$-$s$ junction.  Because the
single-electron charge $e$ is a half of the Cooper-pair charge $2e$,
the frequency of the ac Josephson effect in the $p_x$-$p_x$ junction
is $eV/\hbar$, a half of the conventional Josephson frequency
$2eV/\hbar$ for the $s$-$s$ junction.  These conclusions also apply to
a junction between two cuprate $d$-wave superconductors in such
orientation that both sides of the junction have surface midgap
states, e.g. to the $45^\circ/45^\circ$ junction (see Sec.\ 
\ref{sec:d-wave}).

In both the $p_x$-$p_x$ and $s$-$s$ junctions, electrons transferred
across the interface are taken away into the bulk by the supercurrent
of Cooper pairs.  In the case of the $p_x$-$p_x$ junction, a single
transferred electron occupies a midgap state until another electron
gets transferred.  Then the pair of electrons becomes absorbed into
the bulk condensate, the midgap state returns to the original
configuration, and the cycle repeats.  In the case of the $s$-$s$
junction, two electrons are simultaneously transferred across the
interface and become absorbed into the condensate.  Clearly, electric
charge is transferred across the interface by single electrons at the
rate proportional to $\tau$ in the first case and by Cooper pairs at
the rate proportional to $\tau^2$ in the second case, but the bulk
supercurrent is carried by the Cooper pairs in both cases.

\section{\boldmath Josephson junctions between $d$-wave 
  superconductors}
\label{sec:d-wave}

In this section, we study Josephson junctions between two $d$-wave
cuprate superconductors.  As before, we select the coordinate $x$
perpendicular to the junction line and assume that the electron
momentum component $k_y$ parallel to the junction line is a conserved
good quantum number.  Then, the 2D problem separates into a set of 1D
solutions (\ref{L0s}) in the $x$ direction labeled by the index $k_y$.
Using an isotropic electron energy dispersion law
$\varepsilon=\hbar^2(k_x^2+k_y^2)/2m-\mu$, we replace the Fermi
momentum $k_F$ and velocity $v_F$ by their $x$-components
$k_{Fx}=\sqrt{k_{F}^2-k_y^2}$ and $v_{Fx}=\hbar k_{Fx}/m$.  Thus, the
transmission coefficient D in Eq.\ (\ref{ZDG}) becomes
$k_y$-dependent.  The total Josephson current is given by a sum over
all occupied subgap states labeled by $k_y$.

\begin{figure}[t] 
\includegraphics[width=\linewidth,angle=0]{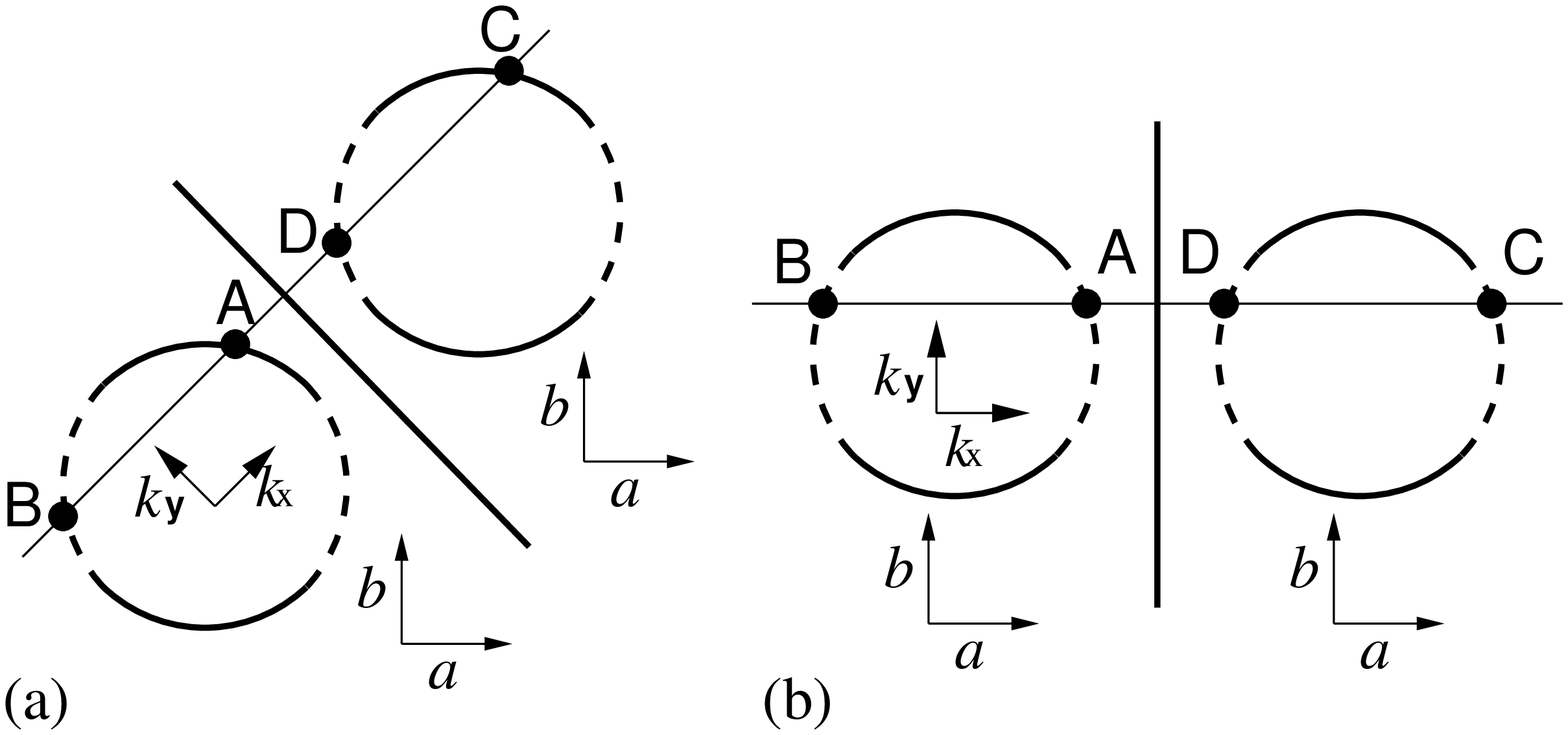} 
\caption{Schematic drawing of the $45^\circ/45^\circ$ junction 
  (panel a) and $0^\circ/0^\circ$ junction (panel b) between two
  $d$-wave superconductors.  The thick line represents the junction
  line.  The circles illustrate the Fermi surfaces, where positive and
  negative pairing potentials $\Delta$ are shown by the solid and
  dotted lines.  The points A, B, C, and D in the momentum space are
  connected by transmission and reflection from the barrier.}
\label{fig:d-wave}
\end{figure}

For the cuprates, let us consider a junction parallel to the
$[1,\bar1]$ crystal direction in the $a$-$b$ plane and select the $x$
axis along the diagonal $[1,1]$, as shown in Fig.\ \ref{fig:d-wave}a.
In these coordinates, the $d$-wave pairing potential is
\begin{equation}
  \hat\Delta_{\sigma k_y}(x,\hat k_x) =
  \sigma2\Delta_\beta\,k_y \hat k_x/k_F^2,
\label{hat-Delta-d}
\end{equation}
where the same notation as in Eq.\ (\ref{hat-Delta}) is used.  Direct
comparison of Eqs.\ (\ref{hat-Delta-d}) and (\ref{hat-Delta})
demonstrates that the $d$-wave superconductor with the $45^\circ$
junction maps to the $p_x$-wave superconductor by the substitution
$\Delta_0\to\sigma2\Delta_0k_y /k_F$.  Thus, the results obtained in
Sec.\ \ref{sec:Q1D} for the $p_x$-$p_x$ junction apply to the
$45^\circ/45^\circ$ junction between two $d$-wave superconductors with
the appropriate integration over $k_y$.  The energies and the wave
functions of the subgap Andreev states in the $45^\circ/45^\circ$
junction are $4\pi$-periodic, as in Eqs.\ (\ref{E_p}).  Thus the ac
Josephson current has the fractional frequency $eV/\hbar$, as in Eq.\ 
(\ref{I_t}).

On the other hand, if the junction is parallel to the $[0,1]$ crystal
direction, as shown in Fig.\ \ref{fig:d-wave}b, then
$\hat\Delta_{\sigma k_y}(x,\hat k_x) = \sigma\Delta_\beta\, (\hat
k_x^2-k_y^2)/k_F^2$.  This pairing potential is an even function of
$\hat k_x$, thus it is analogous to the $s$-wave pairing potential in
Eq.\ (\ref{hat-Delta}).  Thus, the $0^\circ/0^\circ$ junction between
two $d$-wave superconductors is analogous to the $s$-$s$ junction.  It
should exhibit the conventional $2\pi$-periodic Josephson effect with
the frequency $2eV/\hbar$.

For a generic orientation of the junction line, the $d$-wave pairing
potential acts like $p_x$-wave for some momenta $k_y$ and like
$s$-wave for other $k_y$.  Thus, the total Josephson current is a sum
of the unconventional and conventional terms \cite{2phi}:
\begin{equation}
  I=C_1\sin(\phi/2) + C_2\sin(\phi) + \ldots~,
\label{I-phi/2}  
\end{equation}
with some coefficients $C_1$ and $C_2$.  We expect that both terms in
Eq.\ (\ref{I-phi/2}) are present for any real junction between
$d$-wave superconductors because of imperfections.  However, the ratio
$C_1/C_2$ should be maximal for the junction shown in Fig.\ 
\ref{fig:d-wave}a and minimal for the junction shown in Fig.\ 
\ref{fig:d-wave}b.

\section{Experimental observation of the fractional ac Josephson effect}
\label{sec:experiment}

Conceptually, the setup for experimental observation of the fractional
ac Josephson effect is straightforward.  One should apply a dc voltage
$V$ to the junction and measure frequency spectrum of microwave
radiation from the junction, expecting to detect a peak at the
fractional frequency $eV/\hbar$.  To observe the fractional ac
Josephson effect predicted in this paper, it is necessary to perform
this experiment with the $45^\circ/45^\circ$ cuprate junctions shown
in Fig.\ \ref{fig:d-wave}(a).  For control purposes, it is also
desirable to measure frequency spectrum for the $0^\circ/0^\circ$
junction shown in Fig.\ \ref{fig:d-wave}(b), where a peak at the
frequency $eV/\hbar$ should be minimal.  It should be absent
completely in a conventional $s$-$s$ junction, unless the junction
enters a chaotic regime with period doubling \cite{Miracky}.  The
high-$T_c$ junctions of the required geometry can be manufactured
using the step-edge technique.  Bicrystal junctions are not
appropriate, because the crystal axes $\bm{a}$ and $\bm{b}$ of the two
superconductors are rotated relative to each other in such junctions.
As shown in Fig.\ \ref{fig:d-wave}(a), we need the junction where the
crystal axes of the two superconductors have the same orientation.
Unfortunately, attempts to manufacture Josephson junctions from the
Q1D organic superconductors $\rm(TMTSF)_2X$ failed thus far.

The most common way of studying the ac Josephson effect is observation
of the Shapiro steps \cite{Shapiro}.  In this setup, the Josephson
junction is irradiated by microwaves with the frequency $\omega$, and
steps in dc current are detected at the dc voltages
$V_n=n\hbar\omega/2e$.  Unfortunately, this method is not very useful
to study the effect that we predict.  Indeed, our results are
effectively obtained by the substitution $2e\to e$.  Thus, we expect
to see the Shapiro steps at the voltages
$V_m=m\hbar\omega/e=2m\hbar\omega/2e$, i.e.\ we expect to see only
\emph{even} Shapiro steps.  However, when both terms are present in
Eq.\ (\ref{I-phi/2}), they produce both even and odd Shapiro steps, so
it would be difficult to differentiate the novel effect from the
conventional Shapiro effect.  Notice also that the so-called
fractional Shapiro steps observed at the voltage
$V_{1/2}=\hbar\omega/4e$ corresponding to $n=1/2$ have nothing to do
with the effect that we propose.  They originate from the higher
harmonics in the current-phase relation $I\propto\sin(2\phi)$.  The
fractional Shapiro steps have been observed in cuprates \cite{Early},
but also in conventional $s$-wave superconductors \cite{Clarke}.
Another method of measuring the current-phase relation in cuprates was
employed in Ref.\ \cite{ZK}, but connection with our theoretical
results is not clear at the moment.

\section{Conclusions}

In this paper, we study suitably oriented $p_x$-$p_x$ or $d$-$d$
Josephson junctions, where the superconductors on both sides of the
junction originally have the surface Andreev midgap states.  In such
junctions, the Josephson current $I$, carried by the hybridized subgap
Andreev bound states, is a $4\pi$-periodic function of the phase
difference $\phi$: $I\propto\sin(\phi/2)$, in agreement with Ref.\ 
\cite{Kitaev}.  Thus, the ac Josephson current should exhibit the
fractional frequency $eV/\hbar$, a half of the conventional Josephson
frequency $2eV/\hbar$.  In the tunneling limit, the Josephson current
is proportional to the first power of the electron tunneling
amplitude, not the square as in the conventional case
\cite{Tanaka2,Riedel,Barash}.  Thus, the Josephson current in the
considered case is carried by single electrons with charge $e$, rather
than by Copper pairs with charge $2e$.  The fractional ac Josephson
effect can be observed experimentally by measuring frequency spectrum
of microwave radiation from the junction and detecting a peak at
$eV/\hbar$.

The work was supported by NSF Grant DMR-0137726.  KS thanks
S.~M.~Girvin for support.



\begin{thebibliography}{00}
\bibitem{2phi} The current-phase relation $I\propto\sin(\phi/2)$ that
  we propose should not be confused with another unconventional
  current-phase relation $I\propto\sin(2\phi)$ with the period $\pi$,
  which was predicted theoretically for junctions between $s$- and
  $d$-wave superconductors \cite{Yip,Tanaka-sd}, $s$- and $p$-wave
  superconductors \cite{Yip,Tanaka-sp}, and for the $0^\circ/45^\circ$
  junctions between two $d$-wave superconductors \cite{Wendin-ac}.
\bibitem{A-B} Eq.\ (\ref{eq:I_a}) can be equivalently obtained as a
  difference between the right- and left-moving currents using the
  formula $I_a=ev_F(|A_a|^2-|B_a|^2)$ with the normalization condition
  $\int dx\,|\psi^2|=1$ \cite{Furusaki99}.
\end{thebibliography}
\end{document}